\documentclass{achemso}
\usepackage[version=3]{mhchem} 
\usepackage[T1]{fontenc}
\usepackage{lmodern}
\usepackage{amsmath}
\usepackage{bbm}
\usepackage{amsfonts}
\usepackage{multirow}
\usepackage{booktabs}
\usepackage{bm}       
\usepackage{dcolumn} 
\usepackage{titlesec}
\usepackage[round-mode=places,round-precision=2]{siunitx}
\usepackage{blkarray}
\usepackage{color}
\usepackage{float}
\usepackage{graphicx}
\usepackage{subcaption}
\usepackage{caption}
\usepackage{mwe}   
\usepackage{soul}
\usepackage{todonotes} 
\usepackage[integrals]{wasysym}

\setkeys{acs}{articletitle = true}
\setkeys{acs}{chaptertitle = true}
\setkeys{acs}{abbreviations = false}
\setkeys{acs}{maxauthors = 50}
\setkeys{acs}{doi = true}

\newcommand{\lint}{\raisebox{-0.75ex}{\scalebox{1}[2.5]{\(\int\)}}}

\author{Jacob Pedersen$^{\ddagger}$}
\affiliation{Department of Chemistry, Technical University of Denmark, DK-2800 Kongens Lyngby, Denmark}

\author{Bendik Støa Sannes}
\affiliation{Department of Chemistry, Norwegian University of Science and Technology, N-7491 Trondheim, Norway}

\author{Ida-Marie Høyvik}
\email{ida-marie.hoyvik@ntnu.no}
\affiliation{Department of Chemistry, Norwegian University of Science and Technology, N-7491 Trondheim, Norway}

\title{Quantum Statistical Mechanics of Electronically Open Molecules: Reduced Density Operators}

\newcommand{\cre}[1]{a^\dagger_{#1}}
\newcommand{\ann}[1]{a_{#1}}

\newcommand{\anticom}[2]{[#1 , #2]_{+}}
\newcommand{\com}[2]{[#1 , #2]}

\newcommand{\bra}[1]{\langle #1|}
\newcommand{\ket}[1]{| #1 \rangle}

\newcommand{\Hs}{\hat{H}_{\mathrm{M}}}
\newcommand{\He}{\hat{H}_{\mathrm{E}}}

\newcommand{\vack}{\ket{\mathrm{vac}}}
\newcommand{\vacb}{\bra{\mathrm{vac}}}

\begin{document}
\maketitle

\begin{abstract}
    \noindent We present a reduced density operator for electronically open molecules by explicitly averaging over the environmental degrees of freedom of the composite Hamiltonian. Specifically, we include the particle-number non-conserving (particle-breaking) interactions responsible for the sharing of electrons between the molecule and the environment, which are neglected in standard formulations of quantum statistical mechanics. We propose an unambiguous definition of the partial trace operation in the composite fermionic Fock space based on composite states in a second quantization framework built from a common orthonormal set of orbitals. Thereby, we resolve the fermionic partial trace ambiguity. The common orbital basis is constructed by spatial localization of the full orbital space, in which the full composite Hamiltonian naturally partitions into a molecule Hamiltonian, an environment Hamiltonian, and an interaction Hamiltonian. The new reduced density operator is based on the approximation of commutativity between the subsystem Hamiltonians (i.e., molecule and environment Hamiltonians) and the interaction Hamiltonian, which we show corresponds to excluding certain electron transfer channels and neglecting electron transfer relaxation effects. The reduced density operator can be viewed as a generalization of the grand canonical density operator. We are prompted to define the generalized chemical potential, which aligns with the standard interpretation of the chemical potential, apart from the possibility of fractional rather than strictly integer electron transfer in our framework. In contrast to standard approaches, our framework enables an explicit consideration of the electron occupancy in the environment at any level of theory, irrespective of the model used to describe the molecule. Specifically, our reduced density operator is fully compatible with all possible level-of-theory treatments of the environment. The approximations that render our reduced density operator identical to the grand canonical density operator are (i) restriction of excitations to occur within the same orbitals and (ii) assumption of equal interaction with the environment for all molecule spin orbitals (i.e., the wide-band approximation).
\end{abstract}

\newpage
\section{Introduction}

An open quantum system exchanges information, that being energy and/or particles, with its environment,\cite{breuer2002theory,nitzan2006chemical,may2023charge,mukamel1999principles} and we advocate viewing molecules as open quantum systems. Meanwhile, we are not interested in explicitly describing the composite system (molecule and environment). Instead, we pursue reduced descriptions aiming at explicitly treating the molecule with the environmental effects included implicitly. In such reduced description perspectives, information about the molecule will be irreversibly lost to the environment, and the state of the molecule will be accompanied by some degree of uncertainty. Hence, quantum statistical mechanics becomes mandatory to make predictions about the molecule and its properties.\cite{fano1957description} The standard formulation of quantum statistical mechanics is based on (reduced) density operators derived under the approximation of an additively separable composite Hamiltonian, which means that the interaction between the molecule and the environment has been neglected.\cite{zubarev1973nonequilibrium,balescu1975equilibrium,feynman2018statistical} However, this is a crude approximation, since this includes the terms responsible for the sharing of electrons between the molecule and the environment. Therefore, we will in this work derive a reduced density operator for electronically open molecules, i.e., molecules able to share electrons with the environment, by explicitly averaging out the environmental degrees of freedom of the composite Hamiltonian. This new reduced density operator can be viewed as a generalization of the grand canonical density operator.

In the quantum chemistry community, the standard practice for modeling energetically open molecules, i.e., molecules interacting with the environment through electrostatic interactions, is to incorporate the effect of the environmental interaction into the model of the molecule by means of effective Hamiltonians.\cite{guido2020open,csizi2023universal,sun2016quantum,bakowies1996hybrid,olsen2011molecular,kongsted2003coupled,kongsted2002qm,kongsted2003linear,steindal2011excitation,nielsen2007density,aidas2008linear,olsen2015polarizable,barone2022development,curutchet2009electronic,wesolowski2015frozen,hofener2012calculation,folkestad2024quantum,chulhai2017improved,jacob2014subsystem,lee2019projection,chulhai2018projection,saether2017density,hoyvik2020convergence,giovannini2020energy,myhre2013extended,myhre2014multi,myhre2016multilevel,folkestad2019multilevel,folkestad2021multilevel} The use of such effective Hamiltonians can be viewed as facilitating energy dissipation into the environment during an energy minimization (i.e., energetic equilibration) of the molecule. However, it is not yet standard practice to consider electronically open molecules. In response to that, we have recently presented wave-function-based\cite{matveeva2023particle,nee2024particle,pedersen2025time} and density-operator-based\cite{sannes2025fractional} formalisms targeting electronically open molecules by combining standard electronic structure theory\cite{MEST} with open quantum system theory\cite{breuer2002theory,nitzan2006chemical,may2023charge,mukamel1999principles}. Although our approaches are developed to allow molecules to be electronically open with respect to their environment, there is a direct connection to electron transfer theory. In the seminal Marcus electron transfer theory,\cite{marcus1956theory} the electron transfer rates are related to the electronic coupling elements between the donor and acceptor.\cite{marcus1985electron,hush1961adiabatic,hush1968homogeneous} Central to the calculation of these electronic coupling elements is the Hamiltonian matrix element which couples two diabatic (or rather, quasi-diabatic) states.\cite{subotnik2008constructing,blumberger2015recent,hsu1997sequential,kurnikov1996ab} In the so-called direct methods, these coupling elements are evaluated using states obtained by diabatization of the corresponding adiabatic states\cite{pacher1988approximately,cave1996generalization,cave1997calculation,valeev2006effect,biancardi2017electronic,biancardi2016evaluation,biancardi2018benchmark} or computed directly, e.g., from constrained density functional theory.\cite{wu2006extracting,van2010diabatic,oberhofer2010electronic,oberhofer2009charge,de2010derivation} The concepts used to develop the theory for electronically open molecules may similarly be used to calculate such coupling elements since the formalism is based on the construction of quasi-diabatic states for the molecule. Hence, for the application in electron transfer theory, one may construct quasi-diabatic states for the donor and the acceptor, rather than considering a molecule and the environment. This is discussed in more detail by Folkestad and Høyvik in Ref.~\citenum{folkestad2025charge}, where they present electronic coupling elements computed using charge-localized electronic states. The many-electron basis consisting of these charge-localized states enables the formal integration of electronic structure theory with electron transfer theory. In this work, we show that they may similarly be used to formulate quantum statistical mechanics in terms of standard electronic structure theory.

The density operator provides the natural framework for quantum statistical mechanics. Specifically, the uncertainty in a molecular system can be implanted in its state description as a statistical mixture (i.e., probabilistically weighted combination) of maximum-possible-information states (typically called pure states), representing the members of the relevant ensemble.\cite{fano1957description,blum2012density} It is customary to distinguish between equilibrium and non-equilibrium quantum statistical mechanics. In equilibrium theory, the weights of the ensemble states are obtained through a maximization of the information entropy subject to constraints related to the information being exchanged with the surrounding environment.\cite{zubarev1973nonequilibrium,balescu1975equilibrium,feynman2018statistical} The canonical ensemble permits interfacial energy exchange provided that the average energy of the system stays constant. Analogously, the grand canonical ensemble permits interfacial energy and particle fluctuations given that the average energy and average particle number in the system remain constant.\cite{zubarev1973nonequilibrium,balescu1975equilibrium,feynman2018statistical} 

The vantage point for our derivation of the reduced density operator will be the canonical density operator using the composite Hamiltonian. Specifically, we include the interaction term in the Hamiltonian, which is typically neglected. In order to average out the environmental degrees of freedom, we first have to define the partial trace operation in the composite fermionic Fock space. However, this is a non-trivial task due to the correlation between the molecule's and the environment's electronic degrees of freedom, which makes the composite density operator non-separable and the partial trace operation ambiguous. This is known as the fermionic partial trace ambiguity.\cite{friis2013fermionic,montero2011fermionic} We resolve this ambiguity by working with composite states in a second quantization framework based on a common orthonormal set of orbitals, which allows us to propose an unambiguous definition of the partial trace operation in the composite fermionic Fock space. 

In a local spin orbital basis, where the full spin orbital space has been spatially localized onto either the molecule or the environment, the full composite Hamiltonian naturally partitions into a molecule Hamiltonian, an environment Hamiltonian, and an interaction Hamiltonian, which is the structure normally encountered in open quantum system theory.\cite{breuer2002theory,nitzan2006chemical,may2023charge,mukamel1999principles} Neither the molecule nor the environment Hamiltonian commutes with the interaction Hamiltonian. Therefore, the expansion of the canonical composite density operator will follow from the Zassenhaus expansion, that being an infinite product of exponentiated and increasingly nested commutators.\cite{magnus1954exponential,casas2012efficient} By approximating commutativity between the subsystem Hamiltonians and the interaction Hamiltonian, all commutators in the Zassenhaus expansion vanish. Thereby, the canonical density operator gets effectively partitioned into the components of the open quantum system Hamiltonian. We show that by adopting this approximation, we exclude certain electron transfer channels and neglect electron transfer relaxation effects, i.e., the molecule and environment are not allowed to adjust to the altered electron densities after the electron transfer. 

By comparing our reduced density operator to the standard grand canonical density operator, we are prompted to define the generalized chemical potential. The naming of this quantity is motivated by its analogous role and its reduction to the chemical potential under the same approximations that render our reduced density operator identical to the grand canonical density operator. These approximations are (i) restriction of excitations to occur within the same orbitals and (ii) assumption of equal interaction with the environment for all molecule spin orbitals. The restriction of excitations amounts to reducing all one-electron excitation operators to the corresponding molecule and environment number operators. 

The approximation of equal interaction with the environment for all molecule spin orbitals entails the generalized chemical potential to lose its dependence on specific orbitals in the molecule, and thereby, we recover the standard chemical potential. This approximation corresponds to the wide-band approximation,\cite{wingreen1989inelastic,zheng2007time,zhang2013first,verzijl2013applicability} which is noted to be neither intended nor suitable for molecules. The chemical potential is in standard derivations interpreted as the Lagrange multiplier enforcing the constant average electron number in the molecular system,\cite{zubarev1973nonequilibrium,balescu1975equilibrium,feynman2018statistical} thus offering no clues on its underlying approximation or physical soundness. Our work alleviates this problem and provides the missing insights into the chemical potential. Moreover, the chemical potential is noted to determine the electron occupancy (top-down) in standard approaches, whereas our (bottom-up) approach enables an explicit consideration of the electron occupancy in the environment (at any level of theory, irrespective of the model used to describe the molecule) through our definition of the generalized chemical potential. In standard practices, the orbital occupancy of the environment is modeled by, e.g., Fermi--Dirac statistics,\cite{fermi1926sulla,dirac1926theory} which depends on the chemical potential. Hence, the chemical potential serves as a custom parameter in standard practices, since knowledge of the orbital occupancy of the environment requires an explicit quantum mechanical treatment.

The paper is organized as follows. The theoretical preliminaries for our work are presented in Sec.~\ref{sec:EOQS}. Next, we review how to construct states for the composite system, and use this basis to resolve the fermionic partial trace ambiguity by proposing an unambiguous definition of the partial trace operation in the composite fermionic Fock space in Sec.~\ref{sec:define_states}. The reduced density operator is then derived in Sec.~\ref{sec:GC}. Moreover, we compare it with the grand canonical density operator and discuss the physical significance of the approximations invoked in our work. Lastly, we summarize our findings in Sec.~\ref{sec:summary}.

\section{Theoretical Preliminaries} \label{sec:EOQS}

In this section, we present the theoretical preliminaries for our work. Specifically, we provide a brief overview of quantum statistical mechanics from an electronic structure perspective to introduce the density operator formalism. We then briefly discuss the canonical density operator, since it will be our vantage point for the derivation of the reduced density operator. Lastly, we write up the molecular Hamiltonian in a spatially local spin orbital basis and review how it may be partitioned for two interacting subsystems.

\subsection{Quantum Statistical Mechanics}

We consider an ensemble of identical molecular systems, in which each of the members is described by the time-independent Schrödinger equation,
\begin{equation} \label{eq:general_eigenvalue_energy}
    \hat{H} \ket{\Psi_k} = E_k \ket{\Psi_k} ~,
\end{equation}
where $\hat{H}$ is the standard molecular Hamiltonian (specified in eqn.~\eqref{eq:full_hamiltonian}), $\ket{\Psi_k}$ is the $k$th eigenstate of the molecular system, and $E_k$ is the corresponding energy. The molecular system can be described by the density operator defined as\cite{zubarev1973nonequilibrium,balescu1975equilibrium,feynman2018statistical}
\begin{equation} \label{eq:rho_from_eigenstates}
    \hat{\rho} = \sum_k w_k \ket{\Psi_k}\bra{\Psi_k} ~,
\end{equation}
where $w_k$ are the ensemble weights, i.e., the probabilities of the molecular system being in specific eigenstates. The density operator is a convex combination of the eigenstates of the molecular system, meaning that $\sum_k w_k = 1$ and $w_k \ge 0$. Moreover, we note that eqn.~\eqref{eq:rho_from_eigenstates} implicitly assumes that the molecular system is at statistical equilibrium, which means that the corresponding density operator is a stationary solution to the Liouville von Neumann equation (in atomic units),
\begin{equation}\label{eq:liouville_vn_neumann}
    \frac{\partial}{\partial t} \hat{\rho} = -i[\hat{H},\hat{\rho}] ~.
\end{equation}
In other words, $[\hat{H},\hat{\rho}] = 0$, which implies that $\hat{H}$ and $\hat{\rho}$ have simultaneous eigenfunctions. The expectation value of the general operator $\hat O$ of the molecular system can therefore be computed by\cite{zubarev1973nonequilibrium}
\begin{equation}
\langle \hat O\rangle = \mathrm{Tr}(\hat{\rho} \hat{O}) ~.
\end{equation} 

The ensemble weights from eqn.~\eqref{eq:rho_from_eigenstates} are chosen to maximize the information entropy. This ensures that only the available information about the molecular system is used to assign the ensemble weights. The information entropy can be represented by the entropy operator,
\begin{equation}
    \hat \eta = - \ln \hat \rho ~,
\end{equation}
and the average information entropy can thus be computed by\cite{zubarev1973nonequilibrium}
\begin{equation}
    S = \langle \hat{\eta} \rangle = -\mathrm{Tr} (\hat \rho \ln \hat{\rho}) = - \sum_{k} w_k \ln w_k ~.
\end{equation}

\subsection{The Canonical Density Operator}

A molecular system is rarely alone in space. Therefore, we consider each (large) molecular system as energetically equilibrated (i.e., energetically open to its surroundings and at statistical equilibrium). We can make a molecular system energetically open by allowing for small energy fluctuations between the molecular system and its surroundings, and we can ensure that it remains at statistical equilibrium by enforcing the condition of a constant average energy of the molecular system. This entails a canonical distribution of the molecular systems, which means that the members in the ensemble may have different energies, all within the energy window determined by the size of the allowed energy fluctuations. Maximization of the information entropy over this ensemble results in the ensemble weights $w_k =Z^{-1} \exp(-\beta E_k)$ and thus the canonical density operator,\cite{zubarev1973nonequilibrium,balescu1975equilibrium,feynman2018statistical}
\begin{equation}\label{eq:can_density_operator}
    \hat{\rho} = Z^{-1}e^{-\beta \hat{H}} ~,
\end{equation} 
where $\beta$ is defined as the derivative of the average information entropy of the surroundings at total energy with respect to the energy of the surroundings\cite{zubarev1973nonequilibrium,valkunas2013molecular} and $Z = \mathrm{Tr} \exp(- \beta E_k)$ is the canonical partition function, which ensures normalization of the canonical density operator. In the macroscopic (thermodynamic) limit, $\beta$ may be interpreted in terms of inverse temperature (i.e., $\beta=(k_{\mathrm{B}}T)^{-1}$ with $k_{\mathrm{B}}$ being the Boltzmann constant) due to the definition of temperature,\cite{zubarev1973nonequilibrium} but we note that there is no need (nor would it be physically meaningful) to invoke the concept of temperature in the statistical treatment from the electronic-structure perspective.

\begin{figure}[htp!]
    \centering
    \includegraphics[scale=1.55]{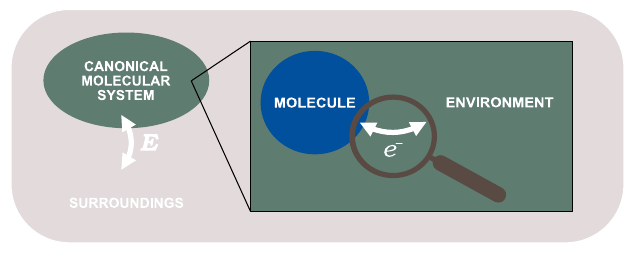}
    \caption{Illustration of our partitioning of a canonical molecular system energetically equilibrated with its surroundings into two interacting subsystems, namely, the molecule and the environment, which can exchange electrons.}
    \label{fig:setup}
\end{figure}

The ensemble weights and therefore the form of the canonical density operator in eqn.~\eqref{eq:can_density_operator} rely on the underlying approximation of an additively separable Hamiltonian (i.e., the interaction between the molecular system and its surroundings is neglected).\cite{zubarev1973nonequilibrium,balescu1975equilibrium,feynman2018statistical} However, we are interested in electronically open molecules, for which we cannot justify such an approximation. As previously mentioned, our vantage point will be a canonically distributed molecular system (and thus the canonical density operator), which we now partition into two interacting subsystems, namely, a small region of interest, henceforth referred to as the molecule, and a large environment surrounding the molecule. The situation is depicted in Fig.~\ref{fig:setup}. The approximation of an additively separable Hamiltonian for the energetically equilibrated molecular system and its surroundings can be justified from our viewpoint (the molecule) because of the non-locality of the neglected interactions (i.e., the neglected interactions between the molecular system and its surroundings occur far away from the molecule). Hence, our vantage point is physically sound and valid for the purpose of this work.

\subsection{Partitioning the Molecular Hamiltonian} \label{sec:partition_H}

For a molecular system at a parametrically fixed nuclear geometry (in the Born--Oppenheimer approximation), the Hamiltonian in eqn.~\eqref{eq:general_eigenvalue_energy} is the standard molecular Hamiltonian. In the spin orbital basis, it takes the following form (excluding nuclear repulsion) in second quantization,\cite{MEST}
\begin{equation}\label{eq:full_hamiltonian}
    \hat{H} = \sum_{PQ} h_{PQ} ~\cre{P} \ann{Q} + \frac{1}{2} \sum_{PQRS} g_{PQRS}~ \cre{P} \cre{R} \ann{S} \ann{Q} ~.
\end{equation} 
The one- and two-electron integrals are defined as
\begin{equation}
    h_{PQ} = \lint \phi^{*}_{P}(\mathbf{x}_{1}) \bigg[ - \frac{1}{2} \nabla^{2} - \sum_{K} \frac{Z_{K}}{|\mathbf{r}_{1}-\mathbf{R}_{K}|} \bigg] \phi_{Q}(\mathbf{x}_{1}) \mathrm{d}\mathbf{x}_{1}
\end{equation}
\begin{equation}
    g_{PQRS} = \lint \lint \phi^{*}_{P}(\mathbf{x}_{1}) \phi_{Q}(\mathbf{x}_{1}) \frac{1}{|\mathbf{r}_{1}-\mathbf{r}_{2}|} \phi^{*}_{R}(\mathbf{x}_{2}) \phi_{S}(\mathbf{x}_{2}) \mathrm{d}\mathbf{x}_{1} \mathrm{d}\mathbf{x}_{2} ~,
\end{equation}
where $\mathbf{x}_{n}$ is the combined spin and spatial $\mathbf{r}_{n}$ coordinate of the $n$th electron, $\mathbf{R}_{K}$ represents the position vector of the $K$th nucleus with nuclear charge $Z_{K}$, and $\{ \phi_{P} \}$ are the spin orbitals. The operators $\cre{P} \ann{Q}$ and $\cre{P} \cre{R} \ann{S} \ann{Q}$ are the one- and two-electron excitation operators, and $\cre{P}$ and $\ann{Q}$ are the fermionic creation and annihilation operators, respectively. The creation and annihilation operators obey the anti-commutation rules for fermionic operators,
\begin{equation} \label{eq:anticomm1}
    \anticom{\cre{P}}{\cre{Q}} = \anticom{\ann{P}}{\ann{Q}} = 0
\end{equation}
\begin{equation}\label{eq:anticomm2}
    \anticom{\ann{P}}{\cre{Q}} = \delta_{PQ} ~.
\end{equation}
 
The spin orbital basis for the composite system $\{ \phi_P\}$ may now be spatially localized such that spin orbitals are localized to either the molecule or the environment. Hence, in the local representation, we may write 
\begin{equation} \label{eq:orbitalsets}
    \{ \phi_P\} = \{\phi_p\} \cup \{\phi_{\bar p}\} ~,
\end{equation}
where $\{\phi_p\}$ and $\{\phi_{\bar p}\}$ denote the sets of spin orbitals localized to the molecule and the environment, respectively. We note that the molecule and environment spin orbitals are mutually orthogonal, since $\{\phi_P\}$ constitutes an orthonormalized set of orbitals. In the absence of covalent molecule-environment bonds, the subsystems have no shared orbital, and the partitioning in eqn.~\eqref{eq:orbitalsets} is therefore possible.\cite{hoyvik2016characterization} In this local spin orbital basis, each term in the Hamiltonian for the composite system (eqn. \eqref{eq:full_hamiltonian}) can be categorized into belonging to the Hamiltonian for the molecule $\hat{H}_{\mathrm{M}}$, the Hamiltonian for the environment $\hat{H}_{\mathrm{E}}$, or the interaction term $\hat{H}_{\mathrm{ME}}$. The full composite Hamiltonian reads
\begin{equation} \label{eq:H}
    \hat{H} =\hat{H}_{\mathrm{M}} + \hat{H}_{\mathrm{E}} + \hat{H}_{\mathrm{ME}} ~.
\end{equation}
The terms in $\hat{H}_{\mathrm{M}}$ and $\hat{H}_{\mathrm{E}}$ include only summation over spin orbitals localized to the molecule and environment, respectively,
\begin{equation}
    \hat{H}_{\mathrm{M}} =\sum_{pq} h_{pq} ~\cre{p}\ann{q} + \frac{1}{2} \sum_{pqrs} g_{pqrs}~ \cre{p} \cre{r} \ann{s} \ann{q}
\end{equation}
\begin{equation}
    \hat{H}_{\mathrm{E}}= \sum_{\bar p \bar q} h_{\bar p \bar q} ~\cre{\bar p}\ann{\bar q} + \frac{1}{2} \sum_{\bar p \bar q \bar r \bar s} g_{\bar p \bar q \bar r \bar s} ~\cre{\bar p} \cre{\bar r} \ann{\bar s} \ann{\bar q} ~.
\end{equation}
The terms in $\hat{H}_{\mathrm{ME}}$ are those which contain mixed summations, and it can be written compactly as
\begin{equation} \label{eq:HME}
 \hat{H}_{\mathrm{ME}} =\hat{H}-\hat{H}_{\mathrm{M}} - \hat{H}_{\mathrm{E}}  ~.
\end{equation}
The interaction Hamiltonian can be split into a particle-number conserving and particle-number non-conserving part. For brevity, these terms are referred to as particle-conserving and particle-breaking, respectively. We refer to our previous work\cite{sannes2025fractional} for the explicit terms and their categorization into particle-conserving and particle-breaking contributions. The particle-conserving interactions are responsible for the interaction between the subsystems and do not cause fluctuations of electrons between the molecule and the environment. These interactions are routinely considered in the quantum chemistry community through embedding and multiscale schemes.\cite{guido2020open,csizi2023universal,sun2016quantum,bakowies1996hybrid,olsen2011molecular,kongsted2003coupled,kongsted2002qm,kongsted2003linear,steindal2011excitation,nielsen2007density,aidas2008linear,olsen2015polarizable,barone2022development,curutchet2009electronic,wesolowski2015frozen,hofener2012calculation,folkestad2024quantum,chulhai2017improved,jacob2014subsystem,lee2019projection,chulhai2018projection,saether2017density,hoyvik2020convergence,giovannini2020energy,myhre2013extended,myhre2014multi,myhre2016multilevel,folkestad2019multilevel,folkestad2021multilevel} The particle-breaking components of $\hat{H}_{\mathrm{ME}}$ can induce fractional charging of the molecule by connecting Fock space states with different particle numbers. The focus of this work is the construction of an electronically open reduced density operator. Therefore, we omit the particle-conserving interactions in the following, but note that they may be included by means of standard practices. We will use the bilinear approximation of the particle-breaking terms in  $\hat{H}_{\mathrm{ME}}$ derived in Ref. \citenum{sannes2025fractional},
\begin{equation} \label{eq:intH}
        \hat{V} = \sum_{p \bar q} ( V_{p \Bar{q}} \cre{p} \ann{\Bar{q}} + V_{\Bar{q} p} \cre{\Bar{q}} \ann{p} ) ~,
\end{equation}
where $V_{p \bar q}$ are the electronic coupling elements representing the strength of the molecule-environment interaction. In the following, we will refer to this effective one-electron particle-breaking interaction Hamiltonian simply as the interaction Hamiltonian. We note that this bilinear form of the interaction Hamiltonian is typically used in descriptions of bipartite open quantum systems.\cite{galperin2012molecular,anderson1961localized,thoss2018perspective,hettler1998nonequilibrium,elenewski2017communication,sun2025control,bergmann2019electron,seshadri2021entropy,mukamel2019flux,cabra2020optical,seshadri2024liouvillian,storm2019computational,pedersen2022redfield,bruneau2014repeated,rivas2012open} However, our derivation in Ref.~\citenum{sannes2025fractional} provides a prescription for how to compute the strengths $V_{p \bar q}$ in terms of standard one- and two-electron integrals.

Furthermore, we note that there is a direct connection between the presented formalism and electron transfer theory. In electron transfer theory, one has an acceptor and a donor (a redox pair), which are both treated explicitly,\cite{marcus1956theory,marcus1985electron,hush1961adiabatic,hush1968homogeneous} rather than a molecule and an environment. If the redox pair is non-covalently bonded, the composite system may be represented in an orbital basis in which each orbital is spatially localized to either the acceptor or the donor,\cite{hoyvik2016characterization} and an analysis of the Hamiltonian equivalent to the one presented in this section is possible for the redox pair. Hence, the particle-breaking terms of  $\hat{H}_{\mathrm{ME}}$ (in this work approximated by $\hat{V}$) is responsible for the electronic coupling elements used in electron transfer theory. This is discussed in more detail by Folkestad and Høyvik in Ref.~\citenum{folkestad2025charge}, where they use charge-localized electronic wave functions to compute electronic coupling elements.

\section{The Composite Fermionic Fock Space}
\label{sec:define_states}

A prerequisite for our work is the ability to average over the environmental degrees of freedom in the composite fermionic Fock space. However, the requirement of anti-commutation between electrons in the molecule and the environment makes this non-trivial. The composite fermionic Fock space is an antisymmetric tensor product (normally called a wedge product) of the molecule and environment Fock spaces. The structure of the wedge product ensures the anti-commutation between electrons in the molecule and the environment, but by partially tracing out the environmental degrees of freedom of the composite density operator, the information necessary to enforce the global anti-commutation rules (eqns.~\eqref{eq:anticomm1}--\eqref{eq:anticomm2}) is lost. This problem is known as the fermionic partial trace ambiguity.\cite{friis2013fermionic,montero2011fermionic} 

Szalay et al. have recently presented an approach to integrate the anti-commutativity of fermionic modes between disjoint sets into the individual matrix elements of the composite density operator, such that the antisymmetric structure of the composite fermionic space is preserved when partially tracing out the non-interesting subsystems.\cite{szalay2021fermionic} Specifically, they apply a Jordan-Wigner transformation\cite{wigner1928paulische} to effectively strip the electrons of their fermionic behavior. The result is a bosonic-like structure containing the information about the fermionic occupations and matrix element-specific sign factors ensuring the fulfillment of the global anti-commutation rules. The fermionic nature of the resulting reduced density operator may then be restored through the inverse transformation.\cite{szalay2021fermionic}

In the present work, we provide an alternative strategy based on the definition of composite states in a second quantization framework built from a common orthonormal set of orbitals. This setup allows us to unambiguously define the partial trace operation in the composite fermionic Fock space, such that the information required to enforce the global anti-commutation rules is absorbed into the matrix elements of the reduced density operator. Thereby, we resolve the fermionic partial trace ambiguity.

This section is organized as follows. First, we review how to construct a composite basis for the interacting composite system from the eigenstates of the non-interacting Hamiltonians. This composite basis is then used to define the fermionic partial trace operation.

\subsection{States for the Non-Interacting Molecule and Environment} \label{sec:states_nonint}

The isolated molecule and environment Hamiltonians describe the non-interacting molecule and environment, respectively, and their eigenstates can be written as
\begin{equation} \label{eq:sys}
    \hat{H}_{\mathrm{M}} \ket{K} = E_{K} \ket{K}
\end{equation}
\begin{equation} \label{eq:env}
    \hat{H}_{\mathrm{E}} \ket{\bar{K}} = E_{\bar{K}} \ket{\bar{K}} ~,
\end{equation}
with $\ket{K}$ ($\ket{\bar K}$) and $E_{K}$ ($E_{\bar K}$) being an eigenstate and corresponding energy eigenvalue of the molecule (environment), e.g., the full configuration interaction eigenpair.

We may further define the molecule and environment number operators as
\begin{equation} \label{eq:sysnumber}
    \hat{N}_{\mathrm{M}} = \sum_{p} \hat{N}_{p} = \sum_{p} \cre{p} \ann{p}
\end{equation}
\begin{equation} \label{eq:envnumber}
    \hat{N}_{\mathrm{E}} = \sum_{\bar p} \hat{N}_{\bar p} = \sum_{\bar p} \cre{\bar p} \ann{\bar p} ~,
\end{equation}
where $\hat{N}_{p}$ and $\hat{N}_{\bar{p}}$ are the corresponding one-electron number operators, and we note that $\{\phi_{p}\}$ and $\{\phi_{\bar{p}}\}$ are orthonormal subsets of the orthonormal composite basis $\{\phi_{P}\}$ (eqn.~\eqref{eq:orbitalsets}). The total number operator is the sum of the molecule and environment number operators,
\begin{equation}
\hat{N} = \hat{N}_{\mathrm{M}} + \hat{N}_{\mathrm{E}} ~.
\end{equation}
It can be shown that $\hat{H}_{\mathrm{M}}$ and $\hat{H}_{\mathrm{E}}$ commute with their respective number operators,
\begin{equation} \label{eq:comNH}
    \com{\hat{N}_{\mathrm{M}}}{\Hs} = \com{\hat{N}_{\mathrm{E}}}{\He} = 0 ~.
\end{equation}
Hence, all eigenstates $\ket{K}$ and $\ket{\bar K}$ are eigenstates of $\hat{N}_{\mathrm{M}}$ and $\hat{N}_{\mathrm{E}}$, respectively, and the states may be additionally labeled according to the number of electrons in the state,
\begin{equation}\label{eq:particle_number_M}
    \hat{N}_{\mathrm{M}} \ket{K, N_{\mathrm{M}}} = N_{\mathrm{M}} \ket{K, N_{\mathrm{M}}}
\end{equation}
\begin{equation}\label{eq:particle_number_E}
    \hat{N}_{\mathrm{E}} \ket{\bar K, N_{\mathrm{E}}} = N_{\mathrm{E}} \ket{\bar K, N_{\mathrm{E}}} ~.
\end{equation}
In the following, the electron numbers are omitted in the state labels to simplify the notation, but we include eqns.~\eqref{eq:particle_number_M}--\eqref{eq:particle_number_E} in order to (i) highlight that each state in the sets $\{\ket{K}\}$ and $\{\ket{\bar K}\}$ corresponds to a particle number eigenstate, and (ii) emphasize that we in this work consider all Fock states, not just those that belong to a particular number of electrons.

\subsection{Many-Electron Basis for the Composite System} \label{sec:composite_basis}

The molecule and environment Hamiltonians commute ($\com{\Hs}{\He} = 0$); therefore, simultaneous eigenstates may be constructed for $\Hs + \He$ from the sets $\{\ket{K}\}$ and $\{\ket{\bar K}\}$. For that purpose, we define state operator strings for the molecule ($A^{\dagger}_{K}$) and the environment ($A^{\dagger}_{\Bar{K}}$),
\begin{equation} \label{eq:statedef}
    A^{\dagger}_{K} = \prod^{N_{K}}_{i \in \ket{K}} \cre{i} \qquad A^{\dagger}_{\Bar{K}} = \prod^{N_{\Bar{K}}}_{\bar i \in \ket{\bar K}} \cre{\bar i} ~,
\end{equation}
where we use the notation $i \in \ket{K}$ and $\bar i \in \ket{\bar K}$ to denote the spin orbitals occupied in the particular molecule and environment state, respectively. The composite states may be defined by means of the state operator strings as\cite{sannes2025fractional}
\begin{equation} \label{eq:compositestate}
    \ket{K \bar{K}} = A^{\dagger}_{K} A^{\dagger}_{\Bar{K}}\ket{\mathrm{vac}} ~,
\end{equation}
with the adjoint of the state defined by
\begin{equation} \label{eq:bracompositestate}
    \bra{K \bar K} \equiv (\ket{K \bar K})^{\dagger} = \vacb A_{\bar K} A_{K} ~.
\end{equation}
The order in which the molecule and environment state operator strings act during the preparation of the composite state is one choice, and we note that all other arrangements would have been equally valid. This leads to a sign ambiguity (called the fermionic ambiguity) in composite states where only one of the subsystems contains an odd number of electrons because of the anti-commutation rules in eqn.~\eqref{eq:anticomm1}.\cite{friis2013fermionic,montero2011fermionic} However, the composite density operator is unaffected by the fermionic ambiguity, since its construction eliminates all potential sign ambiguities. Moreover, we note that the orbitals from which $A_{K}$ and $A_{\bar K}$ are constructed belong to the overall orthonormal set in eqn.~\eqref{eq:orbitalsets}. Therefore, the set $\{\ket{K \bar K}\}$ defines an orthonormal many-electron basis. The composite states $ \ket{K \bar{K}}$ are eigenstates of $\Hs + \He$,
\begin{equation}
(\Hs + \He) \ket{K \bar K} = (E_K + E_{\bar K})\ket{K \bar K} ~, 
\end{equation}
but they are not eigenstates of the full or approximate composite Hamiltonian ($\Hs + \He +\hat{H}_\mathrm{ME}$ or $\Hs + \He +\hat{V}$). The wave function of the composite system may be expanded in this many-electron basis. In this work, the composite wave function will not be constructed explicitly, but its existence will be used in the derivation of the reduced density operator.

\subsection{The Fermionic Partial Trace Operation} \label{sec:partialtrace}

The density operator for the composite system may be written in the basis of the composite states defined in eqns. (\ref{eq:compositestate})--(\ref{eq:bracompositestate}),\cite{blum2012density}
\begin{equation} \label{eq:general_density_operator}
    \hat \rho = \sum_{K L \Bar{K} \Bar{L}}  \rho_{K \bar K, L \bar L}\ket{K \Bar{K}} \bra{L \Bar{L}} ~.
\end{equation}
The matrix elements $\rho_{K \bar K, L \bar L}$ are non-separable due to the correlation between the electronic degrees of freedom in the molecule and the fermionic environment. In other words, the composite density operator cannot be written as a normal tensor product of fermionic density operators representing the molecule and the environment.  \cite{friis2013fermionic,montero2011fermionic,li2001entanglement}

Nonetheless, the troublesome matrix elements are typically approximated as being separable such that the reduced density operator of the molecule can be obtained from the composite density operator by averaging over (i.e., tracing out) the environmental degrees of freedom.\cite{sannes2025fractional,manzano2020short,tupkary2022fundamental,campaioli2024quantum,dzhioev2011super,caban2005entanglement,vidal2002computable,wiseman2003entanglement,ghirardi2004general} The advantage of our composite basis (Sec.~\ref{sec:composite_basis}) is that we can unambiguously define the partial trace operation. Specifically, we propose the following definition of the partial trace operation in the composite fermionic Fock space,
\begin{align} \label{eq:partrace_general}
    \mathrm{Tr}_{\mathrm{E}} ( \hat O ) &= \sum_{K \Bar{K} L} \ket{K} \bra{K \Bar{K}} \hat O \ket{L \Bar{K}} \bra{L}  ~.
\end{align}
We note that this definition can similarly be applied without loss of generality to operators defined on composite bosonic-bosonic and fermionic-bosonic tensor product spaces. 

Our partial trace operation may now be used to trace out the environment states of the composite density operator in eqn.~\eqref{eq:general_density_operator}. The result is the reduced density operator for the molecule,
\begin{align} \label{eq:partrace}
    \hat \sigma = \mathrm{Tr}_{\mathrm{E}} ( \hat \rho ) &= \sum_{K \Bar{K} L} \ket{K} \bra{K \Bar{K}} \hat \rho \ket{L \Bar{K}} \bra{L} = \sum_{K L} \sigma_{KL} \ket{K} \bra{L} ~,
\end{align}
with
\begin{equation} \label{eq:redmatelement}
    \sigma_{KL} = \sum_{\bar K} \rho_{K \bar K, L \bar K} ~.
\end{equation} 
The problem of decomposing the non-separable matrix elements $\rho_{K \bar K, L \bar K}$, such that the antisymmetric structure of the composite space is preserved, has just been turned into how one can construct and compute the reduced matrix elements in eqn.~\eqref{eq:redmatelement}. We can build and compute (or rather, approximate) such matrix elements using our partitioning of the Hamiltonian (Sec.~\ref{sec:partition_H}) and definition of composite states (Sec.~\ref{sec:composite_basis}) combined with the framework of standard electronic structure theory. As previously mentioned, the fermionic partial trace ambiguity arises when the information required to enforce the anti-commutation rules between electrons in the molecule and the environment is lost. This information is contained in the reduced density matrix elements. Consequently, we argue that our definition of the partial trace operation (eqn.~\eqref{eq:partrace_general}) together with our definition of composite states (Sec.~\ref{sec:composite_basis}) resolves the fermionic partial trace ambiguity.

\subsubsection{A Brief Note on Covalent Interactions}

The existence of the spatially localized spin orbital basis in eqn.~\eqref{eq:orbitalsets} is guaranteed as long as the molecule and environment are not connected by a covalent bond (i.e., share orbitals). Therefore, our previous work\cite{matveeva2023particle,nee2024particle,sannes2025fractional,pedersen2025time} has focused on describing non-covalent equilibrium interactions, e.g., molecules interacting with solvent molecules or physisorbing molecular surfaces, but we plan to extend our formalism to covalent bonding situations in future work. However, we note that the presence of covalent bonds will introduce another ambiguity in the reduced description of the molecule related to the partial trace operation. In the following, we comment on this ambiguity in order to (i) avoid confusion with the fermionic partial trace ambiguity (i.e., loss of anti-commutativity between electrons in the molecule and the environment), and (ii) emphasize that we in this work completely avoid this ambiguity by requiring the interaction between the molecule and the environment to be non-covalent. 

The ambiguity arises when distinct composite states give rise to the same reduced description.\cite{friis2013fermionic,wiseman2003entanglement} For example, a bonding interaction between the molecule and an environment may be described as a linear combination of Slater determinants built from orbitals which are strictly localized to the molecule or environment,
\begin{equation}
    \ket{\pm} = \frac{1}{\sqrt{2}} (\cre{p} \pm \cre{\bar{p}}) \ket{M\bar{M}} = \frac{1}{\sqrt{2}} (\cre{p} \pm \cre{\bar{p}}) A^{\dagger}_{M} A^{\dagger}_{\bar{M}} \ket{\mathrm{vac}} ~.
\end{equation}
The symmetric and antisymmetric combination represents ground and excited states. The composite wave function is a pure state, and the composite density operators can be written as
\begin{align}
    \begin{split}
        \hat{\rho}_{\pm} &= \frac{1}{2} (A^{\dagger}_{(M+p)} A^{\dagger}_{\bar{M}} \ket{\mathrm{vac}}\bra{\mathrm{vac}} A_{\bar{M}} A_{(M+p)} \pm (-1)^{N_{M}} A^{\dagger}_{(M+p)} A^{\dagger}_{\bar{M}} \ket{\mathrm{vac}}\bra{\mathrm{vac}} A_{(\bar{M}+\bar{p})} A_{M} \\
        &\pm (-1)^{N_{M}} A^{\dagger}_{M} A^{\dagger}_{(\bar{M}+\bar{p})} \ket{\mathrm{vac}}\bra{\mathrm{vac}} A_{\bar{M}} A_{(M+p)} + A^{\dagger}_{M} A^{\dagger}_{(\bar{M}+\bar{p})} \ket{\mathrm{vac}}\bra{\mathrm{vac}} A_{(\bar{M}+\bar{p})} A_{M} ) ~,
    \end{split}
\end{align}
where we have used the notation $A^{\dagger}_{(M+p)} = \cre{p}A^{\dagger}_{M}$ and $A_{(M+p)} = A_{M}\ann{p}$. However, by partially tracing out the environmental degrees of freedom of each composite density operator, we obtain the same reduced description,
\begin{equation} \label{eq:fermionicambiguity}
    \mathrm{Tr}_{\mathrm{E}} ( \hat{\rho}_{\pm} ) = \frac{1}{2} (A^{\dagger}_{M} \ket{\mathrm{vac}}\bra{\mathrm{vac}} A_{M} + A^{\dagger}_{(M+p)} \ket{\mathrm{vac}}\bra{\mathrm{vac}} A_{(M+p)}) ~,
\end{equation}
and the information detailing the character of the state is lost. This creates an ambiguity in the reduced description, and we note that it is not unique for fermionic-fermionic bipartite systems, but also applies to bosonic-bosonic and fermionic-bosonic composite systems.

\section{The Reduced Density Operator} \label{sec:GC}

In this section, we derive the reduced density operator for electronically open molecules by explicitly averaging out the environmental degrees of freedom of the canonical density operator (eqn.~\eqref{eq:can_density_operator}) equipped with the approximate composite Hamiltonian ($\hat{H}_{\mathrm{M}}+\hat{H}_{\mathrm{E}}+\hat{V}$). The derivation is provided in Sec.~\ref{sec:reduced_operator}, and we construct the corresponding reduced partition function in Sec.~\ref{sec:partition_function}. We then identify the approximations that render our reduced density operator identical to the grand canonical density operator in Sec.~\ref{sec:gc}. By comparing our reduced density operator to the grand canonical density operator, we are prompted to define a quantity that we coin the generalized chemical potential, and its physical significance is elaborated upon in Sec.~\ref{sec:chemicalpotential}. Our derivation of the reduced density operator relies on an approximate commutativity between the subsystem Hamiltonians and the interaction Hamiltonian, and we show in Sec.~\ref{sec:zassenhaus} that this approximation amounts to leaving out certain electron transfer channels and neglecting electron transfer relaxation effects in the molecule and the environment.

\subsection{Derivation} \label{sec:reduced_operator}

We now derive the reduced density operator for an electronically open molecule. Our vantage point will be the canonical composite density operator, i.e., the canonical density operator (eqn. \eqref{eq:can_density_operator}) with the approximate ($\hat{H}_{\mathrm{ME}} \rightarrow \hat{V}$) composite Hamiltonian (eqn. \eqref{eq:H}),
\begin{equation}
    \hat{\rho} = Z^{-1}e^{-\beta(\Hs + \He + \hat{V})} ~.
\end{equation}
The expansion of the canonical composite density operator follows from the Zassenhaus expansion,\cite{magnus1954exponential,casas2012efficient}
\begin{align} \label{eq:expZas}
    \begin{split}
        e^{-\beta ( \Hs + \He + V)} &= e^{-\beta \Hs} e^{-\beta \He} e^{-\beta \hat{V}} e^{\frac{\beta^{2}}{2!}\com{\Hs}{\hat{V}}} e^{\frac{\beta^{2}}{2!}\com{\He}{\hat{V}}} e^{- \frac{\beta^{3}}{3!} 2\com{\He}{\com{\Hs}{\hat{V}}} } \\
        &\times e^{- \frac{\beta^{3}}{3!} (2\com{\hat{V}}{\com{\Hs}{\hat{V}}} + \com{\Hs}{\com{\Hs}{\hat{V}}})}  e^{- \frac{\beta^{3}}{3!} (2\com{\hat{V}}{\com{\He}{\hat{V}}} + \com{\He}{\com{\He}{\hat{V}}})} \cdots ~.
    \end{split}
\end{align}
We now approximate commutativity between the subsystem Hamiltonians and the interaction Hamiltonian. In Sec.~\ref{sec:zassenhaus} we show that this amounts to leaving out certain electron transfer channels and neglecting electron transfer relaxation effects in the molecule and the environment. Thereby, all commutators vanish, and the composite density operator is effectively split into the components of the composite Hamiltonian,
\begin{equation} \label{eq:approxdensity}
    \hat{\rho} \simeq Z^{-1} e^{-\beta\Hs} e^{-\beta\He} e^{-\beta \hat{V}} ~,
\end{equation} 
with $Z$ being the approximate canonical partition function whose role is to ensure normalization of our approximate canonical composite density operator,
\begin{equation} \label{eq:compositeZ}
    Z = \mathrm{Tr} ( e^{-\beta\Hs} e^{-\beta\He} e^{-\beta \hat{V}} ) ~.
\end{equation}
To obtain the reduced density operator, we use the definition of the partial trace operation introduced in Sec.~\ref{sec:partialtrace}. Partially tracing out the environmental degrees of freedom of the approximate canonical composite density operator in eqn.~\eqref{eq:approxdensity} yields
\begin{equation} \label{eq:reduced}
    \hat{\sigma} = Z^{-1} \sum_{K \Bar{K} L} e^{-\beta E_{K}} e^{-\beta E_{\Bar{K}}} \ket{K} \bra{K \Bar{K}} e^{-\beta \hat{V}} \ket{L \Bar{K}} \bra{L} ~,
\end{equation}
where we have used that $\ket{L}$ and $\ket{\Bar{K}}$ are the eigenstates of $\hat{H}_{\mathrm{M}}$ and $\hat{H}_{\mathrm{E}}$, respectively. The exponential of the interaction Hamiltonian is expanded in its Taylor series. Hence, the transition moment reads
\begin{equation} \label{eq:transition_moment}
    \bra{K \Bar{K}} e^{-\beta \hat{V}} \ket{L \Bar{K}} = \bra{K \Bar{K}} \big(1 - \beta \hat{V} + \frac{\beta^{2}}{2} \hat{V}^{2} \mp \cdots \big) \ket{L \Bar{K}} ~.
\end{equation}
In the following, we evaluate $\bra{K \Bar{K}} \hat{V} \ket{L \Bar{K}}$ and $\bra{K \Bar{K}} \hat{V}^{2} \ket{L \Bar{K}}$. 
The first transition moment becomes
\begin{equation} \label{eq:firstTM}
    \bra{K \Bar{K}} \hat{V} \ket{L \Bar{K}} = \bra{K \Bar{K}} \sum_{p \bar q} ( V_{p \Bar{q}} \cre{p} \ann{\Bar{q}} + V_{\Bar{q} p} \cre{\Bar{q}} \ann{p} ) \ket{L \Bar{K}} = 0 ~,
\end{equation}
where we have realized that the composite states contain identical environment states and that the interaction Hamiltonian will either remove or add an electron to the environment. The action of the interaction Hamiltonian will thus result in a vanishing overlap between environment states of different particle number symmetry. In other words, there is no first-order contribution to the reduced density operator, which is consistent with typical open quantum system treatments.\cite{breuer2002theory,nitzan2006chemical,may2023charge,mukamel1999principles} The second transition moment reads
\begin{equation} \label{eq:V2}
    \bra{K \Bar{K}} \hat{V}^{2} \ket{L \Bar{K}} = \sum_{p q \Bar{r} \Bar{s}} [ V_{p \Bar{r}} V_{\Bar{s} q} \bra{K \Bar{K}} \cre{p} \ann{\Bar{r}} \cre{\Bar{s}} \ann{q} \ket{L \Bar{K}} + V_{\Bar{r} p} V_{q \Bar{s}} \bra{K \Bar{K}} \cre{\Bar{r}} \ann{p} \cre{q} \ann{\Bar{s}} \ket{L \Bar{K}} ] ~.
\end{equation}
Next, the resulting transition moments in eqn. \eqref{eq:V2} are evaluated. The first one becomes
\begin{align}\label{eq:trans_mom_1}
    \begin{split}
        \bra{K \Bar{K}} \cre{p} \ann{\Bar{r}} \cre{\Bar{s}} \ann{q} \ket{L \Bar{K}} 
        &= \vacb A_{\Bar{K}} A_{K} \cre{p} \ann{q} \ann{\Bar{r}} \cre{\Bar{s}} A^{\dagger}_{L} A^{\dagger}_{\Bar{K}} \vack \\
        &= \vacb A_{K} \cre{p} \ann{q} A^{\dagger}_{L} A_{\Bar{K}} \ann{\Bar{r}} \cre{\Bar{s}} A^{\dagger}_{\Bar{K}} \vack (-1)^{N_{\Bar{K}}N_{K} + N_{\Bar{K}}N_{L}} \\
        &= \sum_{M \bar{M}} \vacb A_{K} \cre{p} \ann{q} A^{\dagger}_{L} \ket{M \bar{M}} \bra{M \bar{M}} A_{\Bar{K}} \ann{\Bar{r}} \cre{\Bar{s}} A^{\dagger}_{\Bar{K}} \vack \\
        &= \vacb A_{K} \cre{p} \ann{q} A^{\dagger}_{L} \vack \vacb A_{\Bar{K}} \ann{\Bar{r}} \cre{\Bar{s}} A^{\dagger}_{\Bar{K}} \vack \\
        &= \bra{K} \cre{p} \ann{q} \ket{L} \bra{\Bar{K}} \ann{\Bar{r}} \cre{\Bar{s}} \ket{\Bar{K}} ~.
    \end{split}
\end{align}
In the first equality, we have used the definition of the composite states (eqn. (\ref{eq:compositestate})) and moved $\ann{q}$ to the left of $\ann{\Bar{r}} \cre{\Bar{s}}$. In the second equality, we have similarly reordered the state operator strings by molecule and environment degrees of freedom. The reordering of the state operator strings results in the overall sign $(-1)^{N_{\Bar{K}}N_{K} + N_{\Bar{K}}N_{L}}$. Meanwhile, we note that state $\ket{L}$ and $\ket{K}$ must contain the same number of electrons (i.e., $N_{K}=N_{L}$), otherwise, the overlap will be zero. The combined sign factor thus reduces to $(-1)^{2N_{\Bar{K}} N_{K}} = 1$. In addition, we have inserted the resolution of identity in the composite Fock space (i.e., $1 = \sum_{M \bar{M}} \ket{M \bar{M}} \bra{M \bar{M}}$) in the third equality. By using the definition of the composite states and noting that $A^{\dagger}_{\bar{M}}$ and $A_{M}$ must both be the identity for the overlap to be non-zero, we obtain the fourth equality. Lastly, we let the state operator strings work on the vacuum states, thereby separating the expectation value into a molecule and environment part. In a similar way, the second transition moment in eqn. (\ref{eq:V2}) evaluates to
\begin{equation} \label{eq:trans_mom_2}
    \bra{K \Bar{K}} \cre{\Bar{r}} \ann{p} \cre{q} \ann{\Bar{s}} \ket{L \Bar{K}} = \bra{\Bar{K}} \cre{\Bar{r}} \ann{\Bar{s}} \ket{\Bar{K}} \bra{K} \ann{p} \cre{q} \ket{L} ~.
\end{equation}

To simplify eqn.~\eqref{eq:reduced} by means of the results of eqns.~\eqref{eq:trans_mom_1}--\eqref{eq:trans_mom_2}, we first define the canonical-like density operator of the environment, 
\begin{equation}\label{eq:can_densityop_env}
    \hat{\rho}_\mathrm{E} = Z_{\mathrm{E}}^{-1}e^{-\beta\hat{H}_\mathrm{E}} ~,
\end{equation}
with the environment partition function,
\begin{equation}\label{eq:envpartfunc}
    Z_{\mathrm{E}} = \sum_{\bar K}e^{-\beta E_{\bar K}} ~.
\end{equation}
Thereby, we may write eqn.~\eqref{eq:reduced} as
\begin{multline}\label{eq:improveddensity_1}
    \hat{\sigma} \simeq Z^{-1}Z_\mathrm{E} \sum_{K \bar{K} L} e^{-\beta E_{K}} \ket{K} ( \delta_{KL} \delta_{\bar{K}\bar{K}}+ \frac{\beta^{2}}{2} \sum_{p q \Bar{r} \Bar{s}} [ V_{p \Bar{r}} V_{\Bar{s} q} \bra{K}\cre{p}\ann{q}\ket{L} \bra{\bar{K}}\hat{\rho}_\mathrm{E} a_{\bar r}a_{\bar s}^\dagger\ket{\bar{K}} \\
    + V_{\Bar{r} p} V_{q \Bar{s}} \bra{K}\ann{p}\cre{q} \ket{L} \bra{\bar{K}}\hat{\rho}_\mathrm{E}a_{\bar r}^\dagger a_{\bar s}\ket{\bar K} ] + \cdots )\bra{L} ~,
\end{multline} 
where we have used eqn.~\eqref{eq:env} to recover the environment Hamiltonian and thereby write the environment density operator (eqn.~\eqref{eq:can_densityop_env}). We now identify the environment average,
\begin{equation}
    \langle \hat{O}\rangle_\mathrm{E} =\mathrm{Tr}(\hat{\rho}_\mathrm{E}\hat{O})=\sum_{\bar K} Z_\mathrm{E}^{-1}\bra{\bar K} e^{-\beta \hat{H}_\mathrm{E}} \hat{O}\ket{\bar K} ~. 
\end{equation}
Thereby, eqn.~\eqref{eq:improveddensity_1} becomes
\begin{multline} \label{eq:improveddensity}
    \hat{\sigma} \simeq Z^{-1}Z_\mathrm{E} \sum_{K L} e^{-\beta E_{K}} \ket{K} 
    ( \delta_{KL} + \frac{\beta^{2}}{2} \sum_{p q \Bar{r} \Bar{s}} [ V_{p \Bar{r}} V_{\Bar{s} q} \bra{K}\cre{p}\ann{q}\ket{L} \langle a_{\bar r}a_{\bar s}^\dagger\rangle_\mathrm{E} \\
    + V_{\Bar{r} p} V_{q \Bar{s}} \bra{K}\ann{p}\cre{q} \ket{L}\langle a_{\bar r}^\dagger a_{\bar s}\rangle_\mathrm{E} ] + \cdots )\bra{L} ~.
\end{multline}

Next, we anti-commute $\ann{p}\cre{q}$ and $\ann{\bar r}\cre{\bar{s}}$, such that we can simplify the reduced density operator expression by defining a generalized chemical potential,
\begin{equation} \label{eq:generalizedmu}
    \Lambda_{pq} = \frac{\beta}{2} \sum_{\bar r \bar s} [ V_{p \bar r} V_{\bar r q}\delta_{\bar r \bar s} - 2\langle a_{\bar r}^\dagger a_{\bar s}\rangle_\mathrm{E} V_{p \bar s} V_{\bar r q} ] ~,
\end{equation}
and the molecule orbital energy shift,
\begin{equation} \label{eq:orbital_energy_shift}
    \Delta_{p} = \frac{\beta}{2} \sum_{\bar r \bar s} V_{\bar r p} V_{p \bar s} \langle a_{\bar r}^\dagger a_{\bar s}\rangle_\mathrm{E} ~.
\end{equation}
The naming of $\Lambda_{pq}$ is elaborated upon in Sec.~\ref{sec:gc}. In terms of the generalized chemical potential (eqn.~\eqref{eq:generalizedmu}) and the molecule orbital energy shift (eqn.~\eqref{eq:orbital_energy_shift}), the reduced density operator reads
\begin{equation} \label{eq:taylor}
    \hat{\sigma} \simeq Z^{-1} Z_{\mathrm{E}} \sum_{K L} e^{-\beta E_{K}} \ket{K} \bra{K} ( 1 + \beta [ \sum_{pq} \Lambda_{pq} \cre{p}\ann{q} + \sum_{p} \Delta_{p} ] + \cdots ) \ket{L}\bra{L} ~.
\end{equation}
The terms between $\bra{K}$ and $\ket{L}$ resemble the first two terms of the Taylor series of an exponential, and we may approximately recollect the terms in an exponential accordingly. The second-order term of this Taylor series expansion will arise from $\bra{K \bar K} \hat{V}^{4} \ket{L \bar K}$, since all odd powers of $\hat V$ result in overlaps between environment states of different particle-number symmetry, which vanish as argued in eqn.~\eqref{eq:firstTM}. However, we show in Sec.~\ref{sec:zassenhaus} that the first non-vanishing correction to our approximation of commutativity between the subsystem Hamiltonians and the interaction Hamiltonian (eqn.~\eqref{eq:approxdensity}) originates from an effective $\hat{V}^{2}$-type of interaction. Therefore, we decide not to consider any higher-order terms of the Taylor series expansion in eqn.~\eqref{eq:taylor} to stay consistent with this approximation.   

In addition to collecting the terms in an exponential, we now use eqn.~\eqref{eq:sys} to recover the molecule Hamiltonian, $\hat{H}_{\mathrm{M}}$, and the definition of the resolution of the identity (in the molecule Fock space) to bring the reduced density operator into the form,
\begin{equation}
    \hat{\sigma} \simeq X^{-1} \exp[-\beta\Hs] \exp [\beta ( \sum_{pq} \Lambda_{pq} \cre{p}\ann{q} + \sum_{p} \Delta_{p} ) ] ~,
\end{equation}
where we have also identified the reduced partition function for the molecule, $X = Z Z^{-1}_{\mathrm{E}}$. This reduced partition function is considered in more detail in Sec.~\ref{sec:partition_function}. We note that the orbital energy shift (defined in eqn.~\eqref{eq:orbital_energy_shift}) is a constant scalar and commutes with all operators. It may therefore be moved to the first exponential and be absorbed into the molecule Hamiltonian, 
\begin{equation} \label{eq:redefineH}
    \Hs \rightarrow \Hs - \sum_{p} \Delta_{p} ~.
\end{equation}
The molecule orbital energy shift directly results from the relaxation induced by the coupling of the molecule to the interacting environment. Consequently, the reduced density operator becomes
\begin{equation} \label{eq:reduced_final}
    \hat{\sigma} \simeq X^{-1} \exp\big[ -\beta \Hs] \exp\big[\beta \sum_{pq} \Lambda_{pq} \cre{p}\ann{q} \big] ~.
\end{equation}

\subsection{The Reduced Partition Function} \label{sec:partition_function}

The reduced partition function $X=Z Z_\mathrm{E}^{-1}$ ensures the normalization of the reduced density operator. The definition of $Z_\mathrm{E}$ is given in eqn.~\eqref{eq:envpartfunc}, and the canonical composite partition function defined in eqn.~(\ref{eq:compositeZ}) can be written as
\begin{equation} \label{eq:comZ}
    Z = \sum_{K \bar K} e^{-\beta E_{K}} e^{-\beta E_{\bar K}} \bra{K \bar K} e^{-\beta \hat{V}} \ket{K \bar{K}} ~,
\end{equation}
where we have exploited that $\ket{K}$ and $\ket{\bar K}$ are eigenstates of $\Hs$ and $\He$, respectively. The expectation value in eqn.~(\ref{eq:comZ}) is evaluated following the same procedure as shown from eqn.~(\ref{eq:transition_moment}) to (\ref{eq:reduced_final}), the only difference being $\ket{K \bar K}$ instead of $\ket{L \bar K}$, which directly entails the overall sign $(-1)^{2 N_{\bar K} N_{K}} = 1$. The reduced partition function therefore becomes
\begin{align}
    \begin{split}
        X &= Z Z^{-1}_{\mathrm{E}} \\
        &= \sum_{K} \bra{K}  \exp[-\beta \Hs]( 1 + \beta \sum_{pq} ( \Lambda_{pq} \cre{p}\ann{q} + \Delta_{p} )+\cdots ) ] \ket{K} \\
        &= \sum_{K} \bra{K} \exp[-\beta \Hs] \exp \big[ \beta  \sum_{pq} \Lambda_{pq} \cre{p}\ann{q}\big]  \ket{K} ~,
    \end{split}
\end{align}
where the last equality follows from absorbing the constant molecule orbital energy shifts into the molecule Hamiltonian, as done in eqn.~\eqref{eq:redefineH}.

\subsection{The Grand Canonical Density Operator} \label{sec:gc}

We now want to connect our reduced density operator (eqn.~\eqref{eq:reduced_final}) to the grand canonical density operator. In other words, we want to identify the approximations that will render our reduced density operator identical to the grand canonical density operator. This will provide insights into the underlying approximations and physical content of the grand canonical density operator and the chemical potential. 

By restricting the excitation operators in eqn.~\eqref{eq:reduced_final} to occur within the same orbitals, the one-electron excitation operators for the molecule and environment reduce to the corresponding molecule and environment one-electron number operators as defined in eqns.~(\ref{eq:sysnumber})--(\ref{eq:envnumber}), respectively. Specifically, the restriction of the environment one-electron excitation operators ($\cre{\bar r}\ann{\bar s} \rightarrow  \cre{\bar r}\ann{\bar r} \delta_{\bar r \bar s} =  \hat{N}_{\bar r}\delta_{\bar r \bar s}$) entails the generalized chemical potential to read
\begin{align} \label{eq:twoelesimpl}
    \begin{split}
        \Lambda_{pq} & \rightarrow \frac{\beta}{2} \sum_{\bar r \bar s} [ V_{p \bar r} V_{\bar r q}\delta_{\bar r \bar s} - 2\langle a_{\bar r}^\dagger a_{\bar r}\rangle_\mathrm{E} \delta_{\bar r \bar s}V_{p \bar s} V_{\bar r q}  ]\\
        &=\frac{\beta}{2} \sum_{\bar r } V_{p \bar r} V_{\bar r q} [1 - 2\langle \hat{N}_{\bar r}\rangle_\mathrm{E} ] ~. 
    \end{split}
\end{align}
The restriction on the molecule one-electron excitation operators ($\cre{p}\ann{ q} \rightarrow  \cre{p}\ann{q} \delta_{pq} = \hat{N}_{p}\delta_{pq}$) results in
\begin{equation}
    \sum_{pq}\Lambda_{pq}\cre{p}\ann{q} \rightarrow \sum_{pq}\Lambda_{pq}\cre{p}\ann{p}\delta_{pq} = \sum_{p}\Lambda_{pp}\hat{N}_p ~. 
\end{equation}
$\Lambda_{pp}$ are the diagonal elements of eqn. \eqref{eq:twoelesimpl}, and can be written as 
\begin{align} \label{eq:genchempotdiag}
    \begin{split}
        \Lambda_{pp}&= \frac{\beta}{2} \sum_{\bar q} V_{p \bar q} V_{\bar q p}[1 - 2\langle \hat{N}_{\bar q}\rangle_\mathrm{E} ] \\
        &\equiv \frac{\beta}{2} \sum_{\bar q}\mu_{p\bar q} ~, 
    \end{split}
\end{align}
where we have introduced the one-electron chemical potential,
\begin{equation}\label{eq:chempot}
    \mu_{p\bar q}= V_{p \bar q} V_{\bar q p}[1 - 2\langle \hat{N}_{\bar q}\rangle_\mathrm{E} ] ~.
\end{equation}

Similarly, under this restriction on the excitation operators, the molecule orbital energy shift $\Delta_{p}$ used in the redefinition of the molecule Hamiltonian becomes
\begin{equation}
    \Delta_{p} \rightarrow  \frac{\beta}{2}  \sum_{\bar q} V_{\bar q p} V_{p \bar q} \langle \hat{N}_{\bar q}\rangle_\mathrm{E} ~.
\end{equation}
Finally, by approximating all molecule spin orbitals to have equal interaction with the environment (i.e., $V_{p\bar q} \rightarrow V_{\bar q}$), the generalized chemical potential is the same for all molecule spin orbitals, and we may write
\begin{equation}
    \Lambda_{pp}\rightarrow \mu ~.
\end{equation} 
This approximation amounts to the wide-band approximation known from non-equilibrium transport theory.\cite{wingreen1989inelastic,zheng2007time,zhang2013first,verzijl2013applicability} Altogether, the approximated reduced density operator becomes
\begin{align} \label{eq:GCdensity}
    \begin{split}
        \hat{\sigma} & \rightarrow \Xi^{-1} \exp[ -\beta \Hs] \exp[\beta \mu\sum_{p}  \hat{N}_p]\\
        &= \Xi^{-1} \exp[ -\beta \Hs] \exp[\beta  \mu \hat{N}_\mathrm{M}]\\
        &= \Xi^{-1}\exp[ -\beta (\Hs-\mu\hat{N}_\mathrm{M}) ] ~,
    \end{split}
\end{align}
where we have also observed that the reduced partition function reduces to the standard grand canonical partition function under these approximations (i.e., $X \rightarrow \Xi$). In addition, we have used that the molecule Hamiltonian and molecule number operator commute (eqn.~\eqref{eq:sysnumber}) to gather the terms in a single exponential in the last equality. In contrast, $[\hat{H}_\mathrm{M}, \cre{p}\ann{q}] \neq 0$, which means that the reduced density operator in eqn.~\eqref{eq:reduced_final} cannot be collected in one exponential. The result of eqn.~\eqref{eq:GCdensity} is the grand canonical density operator.

We note that expectation values of the environment one-electron excitation operators naturally reduce to the environment number operators in the case of a single-determinantal environment wave function. However, this is not the case for the molecule one-electron excitation operators nor if multi-determinantal wave functions, such as coupled cluster and configuration interaction wave functions, are used for describing the environment.

\subsection{The Generalized Chemical Potential} \label{sec:chemicalpotential}

An important feature of our reduced density operator (eqn.~\eqref{eq:reduced_final}) relative to the grand canonical density operator (eqn.~\eqref{eq:GCdensity}) is the independence of the generalized chemical potential with respect to the level of theory used to describe the environment. In other words, our reduced density operator is fully compatible with all possible level-of-theory treatments of the environment, ranging from full configuration interaction theory to Fermi-Dirac statistics.\cite{fermi1926sulla,dirac1926theory} This is in stark contrast to standard protocols, where the chemical potential determines the environment orbital occupancy,\cite{zubarev1973nonequilibrium,balescu1975equilibrium,feynman2018statistical} and not the other way around as in our framework. 

The diagonal values of the generalized chemical potential (eqn.~\eqref{eq:genchempotdiag}) are interpreted as the energy released or absorbed by the molecule upon the removal or addition, respectively, of a (fractional) electron. We denote this process as direct electron transfer. This aligns with the standard interpretation of the chemical potential, apart from the possibility of fractional rather than strictly integer electron transfer in our framework.\cite{zubarev1973nonequilibrium,balescu1975equilibrium,feynman2018statistical} The off-diagonal elements are interpreted as the energy contribution following the coupling to the environment on the excitations within the molecule. This process is termed indirect electron transfer.

\begin{figure}
    \centering
    \includegraphics[scale=1.75]{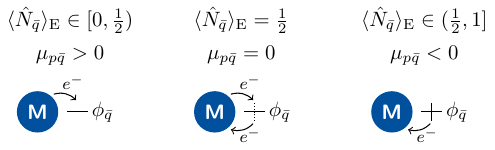}
    \caption{The direction of electron flow in non-equilibrium situations can be predicted by the average environment orbital occupation. The different situations are: favored electron transfer to the environment (left), equilibrium between electron addition and removal (center), and favored electron transfer to the molecule (right). The molecule's ability to donate or accommodate an electron is taken into account by the molecule number operator. Therefore, the direction of electron flow is independent of the specific molecule spin orbitals.}
    \label{fig:chempot}
\end{figure}

Insight into the direction of electron flow in non-equilibrium situations can be obtained from our definition of the generalized chemical potential. Specifically, a positive chemical potential (i.e., energy is released) represents the situation where an electron can be transferred from the molecule to the environment, while a negative chemical potential (i.e., energy is absorbed) corresponds to an electron being transferred from the environment to the molecule. In other words, the direction of the electron flow in non-equilibrium situations is dictated by the sign of the chemical potential, as is standard following the principle of electronegativity equalization\cite{sanderson1951interpretation,rappe1991charge}. The sign of our generalized chemical potential directly follows from the value of $(1 - 2 \langle \hat{N}_{\bar q}\rangle_\mathrm{E})$. The average orbital occupancy of the environment can take any value in the range $\langle \hat{N}_{\bar q}\rangle_\mathrm{E} \in [0, 1]$, which implies that $(1 - 2 \langle \hat{N}_{\bar q}\rangle_\mathrm{E}) \in [-1, 1]$. Consequently, if the environment orbital on average is unfilled or less than half-filled, the generalized chemical potential will be positive, and electron transfer from the molecule to the environment is favored [Fig.~\ref{fig:chempot}(left)]. On the other hand, if the environment orbital on average is more than half-filled or filled, the generalized chemical potential will be negative, and electron transfer from the environment to the molecule is favored [Fig.~\ref{fig:chempot}(right)]. A special situation arises when $\langle \hat{N}_{\bar q}\rangle_\mathrm{E} = 0.5$, entailing $(1 - 2 \langle \hat{N}_{\bar q}\rangle_\mathrm{E}) = 0$, which represents an equilibrium situation between electron addition and removal [Fig.~\ref{fig:chempot}(center)]. This aligns with our expectations, and the derivation reveals that the direction of electron flow is built directly into our definition of the generalized chemical potential.

Lastly, we note that the equation for the diagonal elements of our generalized chemical potential (i.e., direct electron transfer, eqn.~\eqref{eq:genchempotdiag}) resembles the Fermi golden rule\cite{dirac1927quantum} type of equation typically encountered in electron transfer theory.\cite{barbara1996contemporary,fletcher2010theory,di2008electrical} Specifically, it depends on the squared coupling elements and an internal occupation number condition to ensure a physically meaningful transfer process, i.e., whether the environment can donate or accommodate the transferred electron. The molecule's ability to donate or accommodate an electron is taken into account by the molecule number operator, and we note that overall particle conservation is built into the equations. This may be demonstrated by writing the diagonal element of the first non-vanishing coupling element responsible for the electron transfer (eqn.~\eqref{eq:V2}) as
\begin{align} \label{eq:partcons}
    \begin{split}
        \bra{K \Bar{K}} \hat{V}^{2} \ket{K \Bar{K}} 
        &= \sum_{p \Bar{q}} V_{p \Bar{q}} V_{\Bar{q} p} [ \bra{K} \hat{N}_{p} \ket{K} \bra{\Bar{K}} (1-\hat{N}_{\bar{q}}) \ket{\Bar{K}} + \bra{\Bar{K}} \hat{N}_{\Bar{q}} \ket{\Bar{K}} \bra{K} (1-\hat{N}_{p}) \ket{K} ] \\
        &= \sum_{p \Bar{q}} V_{p \Bar{q}} V_{\Bar{q} p} (N^{K}_{p} + N^{\bar{K}}_{\bar{q}} - 2 N^{K}_{p} N^{\bar{K}}_{\bar{q}}) ~,
    \end{split}
\end{align}
where $N^{K}_{p}$ and $N^{\bar{K}}_{\bar{q}}$ denote the electron number in orbital $\phi_{p}$ and $\phi_{\bar{q}}$ in state $\ket{K}$ and $\ket{\bar{K}}$, respectively.
Hence, eqn.~\eqref{eq:partcons} shows that direct electron transfer can only occur if the molecule and environment spin orbital pair constitute one occupied and one virtual spin orbital. This is the particle-number equivalent of the resonance (i.e., energy conservation) condition in Fermi's golden rule.

\subsection{Non-Commutativity Effects} \label{sec:zassenhaus}

In this section, we show that the approximation of commutativity between the subsystem Hamiltonians and the interaction Hamiltonian invoked in eqn. (\ref{eq:approxdensity}) amounts to leaving out certain electron transfer channels and neglecting electron transfer relaxation effects in the molecule and environment. Therefore, we retain the next higher-order terms (with respect to $\beta$) in the Zassenhaus expansion, i.e., the exponentials including the commutators $\com{\Hs}{\hat{V}}$ and $\com{\He}{\hat{V}}$, such that the approximate canonical composite density operator now reads  
\begin{equation}
    \hat{\rho} \simeq Z^{-1} e^{-\beta\Hs} e^{-\beta\He} e^{-\beta \hat{V}} e^{\frac{\beta^{2}}{2!}\com{\Hs}{\hat{V}}} e^{\frac{\beta^{2}}{2!}\com{\He}{\hat{V}}} ~,
\end{equation}
with $Z$ being the corresponding approximate canonical composite partition function. Thereby, the reduced density operator becomes
\begin{equation}
    \hat{\sigma} = Z^{-1} \sum_{\substack{K \Bar{K} L \\ M \bar M N \bar N}} e^{-\beta E_{K}} e^{-\beta E_{\Bar{K}}} \ket{K} \bra{K \Bar{K}} e^{-\beta \hat{V}} \ket{M \bar M} \bra{M \bar M} e^{\frac{\beta^{2}}{2!}\com{\Hs}{\hat{V}}} \ket{N \bar N} \bra{N \bar N} e^{\frac{\beta^{2}}{2!}\com{\He}{\hat{V}}} \ket{L \Bar{K}} \bra{L} ~.
\end{equation}
The exponentials within the transition moments are expanded in their respective Taylor series and regrouped,
\begin{multline}
    \hat{\sigma} = Z^{-1} \sum_{K \Bar{K} L} e^{-\beta E_{K}} e^{-\beta E_{\Bar{K}}} \ket{K}  [ \delta_{KL} - \beta \bra{K \Bar{K}} \hat{V} \ket{L \bar K} + \frac{\beta^{2}}{2} \bra{K \bar K} ( \hat{V}^{2} + \com{\Hs}{\hat{V}} + \com{\He}{\hat{V}} ) \ket{L \bar K} \\
    - \sum_{M \bar M} \frac{\beta^{3}}{2} \bra{K \bar K} \hat{V} \ket{M \bar M} \bra{M \bar M} ( \com{\Hs}{\hat{V}} + \com{\He}{\hat{V}} ) \ket{L \bar K} + \cdots ] \bra{L} ~.
\end{multline}
Since $\hat{H}_\mathrm{M}$ is particle-conserving and $\hat{V}$ is particle-breaking, the action of $\com{\Hs}{\hat{V}}$ and $\com{\He}{\hat{V}}$ will be to effectively move one electron interfacially. This implies that the additional first order contributions, $\bra{K \bar K} \com{\Hs}{\hat{V}} \ket{L \bar K}$ and $\bra{K \bar K} \com{\He}{\hat{V}} \ket{L \bar K}$, to the reduced density operator vanish, following the same argument as in eqn.~\eqref{eq:firstTM}. Hence, the reduced density operator simplifies to
\begin{multline} \label{eq:reducedextra}
    \hat{\sigma} = Z^{-1} \sum_{K \Bar{K} L} e^{-\beta E_{K}} e^{-\beta E_{\Bar{K}}} \ket{K}  [ \delta_{KL} + \frac{\beta^{2}}{2} \bra{K \bar K} \hat{V}^{2} \ket{L \bar K} \\ - \sum_{M \bar M} \frac{\beta^{3}}{2} \bra{K \bar K} \hat{V} \ket{M \bar M} \bra{M \bar M} ( \com{\Hs}{\hat{V}} + \com{\He}{\hat{V}} ) \ket{L \bar K} \pm \cdots ] \bra{L} ~.
\end{multline}
A comparison of eqn.~(\ref{eq:reducedextra}) with eqns.~(\ref{eq:reduced})--(\ref{eq:transition_moment}) reveals that the first non-vanishing correction to the reduced density operator is
\begin{equation} \label{eq:correction}
    \sum_{K \bar K L M \bar M} \frac{\beta^{3}}{2} \bra{K \bar K} \hat{V} \ket{M \bar M} \bra{M \bar M} ( \com{\Hs}{\hat{V}} + \com{\He}{\hat{V}} ) \ket{L \bar K} ~.
\end{equation}
The operation-wise similarity between the interaction Hamiltonian and the commutators $\com{\Hs}{\hat V}$ and $\com{\He}{\hat V}$ (i.e., both work by effectively moving one electron interfacially) means that the correction term in eqn.~(\ref{eq:correction}) facilitates electron transfer through different channels (compared to that induced by $\hat{V}^{2}$). In addition, it encodes electron transfer relaxation effects, i.e., effects stemming from the altered molecule and environment charge densities induced by the electron transfer, as captured by the presence of the subsystem Hamiltonians.

\section{Summary} \label{sec:summary}
  
In this work, we derive a reduced density operator for electronically open molecules by explicitly tracing out the environmental degrees of freedom of the composite Hamiltonian. Specifically, we include the particle-number non-conserving (particle-breaking) interactions responsible for electron fluctuations between the molecule and the environment, which are neglected in standard formulations of quantum statistical mechanics. The derivation is based on composite states in a second quantization framework built from a common orthonormal set of orbitals. This allows us to propose an unambiguous definition of the partial trace operation in the composite fermionic Fock space. Thereby, we resolve the fermionic partial trace ambiguity.

The full composite Hamiltonian naturally splits into a molecule Hamiltonian, an environment Hamiltonian, and an interaction Hamiltonian in a spatially localized spin orbital basis. The new reduced density operator is based on the approximation of commutativity between the subsystem Hamiltonians (i.e., molecule and environment Hamiltonians) and the interaction Hamiltonian. This entails the canonical composite density operator to be effectively partitioned into a molecule, an environment, and an interaction component. By identifying the first correction to this non-commutativity between the subsystem Hamiltonians and interaction Hamiltonian, we show that this approximation amounts to excluding certain electron transfer channels and neglecting electron transfer relaxation effects.

The new reduced density operator can be viewed as a generalization of the grand canonical density operator. Specifically, we identify the approximations that render our reduced density operator equal to the grand canonical density operator. These approximations are (i) restriction of excitations to occur within the same orbitals and (ii) assumption of equal interaction with the environment for all molecule spin orbitals (i.e., the wide-band approximation). This provides new insights into the underlying approximations and physical content of the grand canonical density operator.

We are prompted to define the generalized chemical potential. The naming of this quantity is motivated by its analogous role and its reduction to the standard chemical potential under the approximations that connect our reduced density operator to the grand canonical density operator. The diagonal elements of the generalized chemical potential are interpreted as the energy associated with the addition or removal of a (fractional) electron. This aligns with the standard interpretation of the chemical potential, apart from the possibility of fractional rather than strictly integer electron transfer in our framework. The chemical potential is in standard formulations of quantum statistical mechanics interpreted as the Lagrange multiplier enforcing the constant average electron number in the molecular system. Meanwhile, our framework enables an explicit consideration of the electron occupancy in the environment at any level of theory, irrespective of the model used to describe the molecule. Consequently, our reduced density operator is fully compatible with all possible level-of-theory treatments of the environment.

We obtain insights into the direction of electron flow in non-equilibrium situations from our definition of the generalized chemical potential. We find that if the environment orbital on average is unfilled or less than half-filled, the generalized chemical potential will be positive, and electron transfer from the molecule to the environment is favored. Similarly, if the environment orbital on average is more than half-filled or filled, the chemical potential will be negative, and electron transfer from the environment to the molecule is favored. If the environment orbital is half-filled on average, the chemical potential is zero, and this represents an equilibrium situation between electron addition and removal. The equation for the diagonal elements of the generalized chemical potential (i.e., direct electron transfer) resembles the Fermi golden rule type of equation, and we show that overall particle conservation is built into the equations. This is the particle-number equivalent of the resonance (i.e., energy conservation) condition in Fermi's golden rule.

Lastly, we note that our derivation is based on the assumption of a non-covalent equilibrium interaction between the molecule and the environment. It is of great interest to extend this to include covalent bonding and non-equilibrium situations, and we will in future work explore such extensions.

\section{Acknowledgments}

J.P. acknowledges financial support from the Technical University of Denmark within the Ph.D. Alliance Programme, Knud Højgaards Fond (grant no. 23-02-2571), William Demant Fonden (grant no. 23-3137), and Ingeniør Alexandre Haynman og Hustru Nina Haynmans Fond (grant no. KR-4110605). I-M.H. and B.S.S. acknowledge funding from the Research Council of Norway through FRINATEK project 325574. I-M.H. acknowledges support from the Centre for Advanced Study in Oslo, Norway, which funded and hosted her Young CAS Fellow research project during the academic year of 22/23 and 23/24.

\section{Data Availability Statement}

No data were generated or analyzed in support of this work.

\bibliography{references}

\section{TOC Image}
\begin{center}
    \includegraphics[scale=1.75]{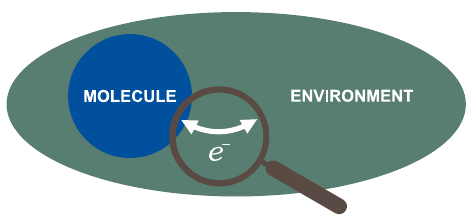}
\end{center}

\end{document}